# Overview of the SiLC R&D activities

Aurore Savoy-Navarro[1*]

1 – LPNHE – Université Pierre et Marie Curie/IN2P3-CNRS
4, Place Jussieu, 75252 Paris-Cedex05 – France

The R&D Collaboration SiLC (Silicon tracking for Linear Colliders) is based on generic R&D aiming to develop the next generation of large Silicon tracking systems for the Linear collider experiments; it serves all three ILC detector concepts. There is a strong involvement in ILD, a natural collaboration with SiD and recent 4[th] concept interest to use Silicon tracking technology as well. Here is a very brief summary of the latest results on sensors, Front End Electronics, Mechanics and Integration issues, test bench and test beam results and where to go from there.

## 1  Introduction

Since 2002, the SiLC R&D worldwide collaboration [1] is addressing the main R&D objectives to develop the next generation of large Silicon tracking systems for the future Linear Collider experiments in the case both of an all-Silicon tracking system (SiD and eventually 4[th] concept) or of a combined Silicon tracking system with a gaseous central tracker (ILC or 4[th] concept). It is a *horizontal R&D* that applies to all three detector concepts. The main R&D lines and presently achieved results are briefly discussed.

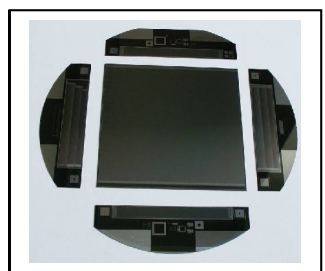

**Figure 1:** HPK strip sensor test structures

## 2  Main R&D objectives and present results

The main R&D objectives are on new sensors, new Front End electronics on detector and related DAQ, new mechanical structures in order to achieve at best the main goals of the next generation Silicon trackers: low material budget, higher tracking performances (with respect even to LEP), robustness, reliability, easy to construct and to calibrate/monitor.

### 2.1  R&D on sensors

The short term baseline for sensors is based on micro-strips sensors. Today technology corresponds to sensors developed for SiLC by Hamamatsu Photonics (HPK), made on 6'' wafers, with 50μm pitch, 320 μm thick and including test structures (Fig.1). Some of them have a special treatment for alignment. They were tested in details including at a test beam [2] For the short term, more performing strip detectors (e.g. edgeless micro-strip sensors as developed by CANBERRA) are considered as well as HPK thinner sensors and novel technology, i.e. edgeless planar or 3D planar, under development by VTT, IRST and CNM.
The longer term objectives include pixel detectors mainly based on 3D technology; for example Low Material Budget and High Intrinsic Gain (LMB) 3D pixels are based on microcells [3]. They are operating at breakdown mode gain up to $10^6$ with a thickness down to 10μm and pixel size from 10x10 μm$^2$ to 100x100 μm$^2$ (Fig.2).

---

[*] On behalf of SiLC R&D Collaboration: the list of Institutes and people collaborating to the SiLC R&D can be found in Reference [1].

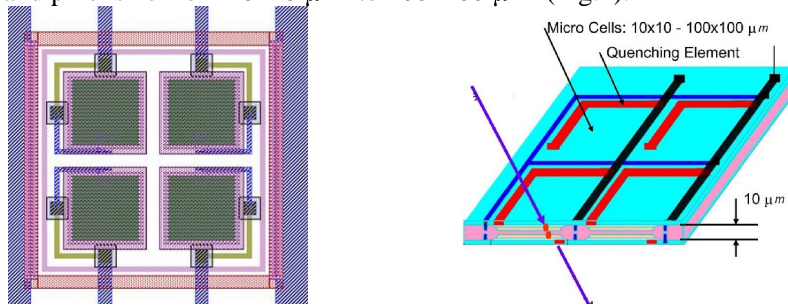

**Figure 2:** Schematic view (left) and principle of functioning (right) of the LMB pixels

## 2.2 R&D on Electronics:

### 2.2.1 *A highly performing and with high degree of signal processing FEE*

A new FE readout ASIC is produced in 130nm CMOS UMC technology with a sophisticated degree of signal processing and a fully programmable operation mode [4]. They will equip the test beam Silicon prototypes in the EUDET E.U. program. The next steps are elementary blocks (chips) for treating 256 channels, chip thinning and going to 90nm CMOS technology.

### 2.2.2 *Direct connection of the FEE onto the detector*

Developing the direct connection of the FE chip onto the strip sensors is a major goal, which is proceeding in two steps, i) first by bump bonding, followed by ii) 3D vertical interconnect.

### 2.2.3 *Starting developing the DAQ chain*

The R&D on the new DAQ system has started well motivated by the test beam needs that serve also as training camps. A major step was achieved in 2008 by developing a fully standalone tracking Silicon system easy to connect to other sub detectors DAQ systems [5].

## 2.3 Mechanics and Detector integration issues

The R&D work on mechanics is focusing on a series of crucial issues such as i) alignment essential for Physics that requires very high spatial resolution [6], ii) new modules and light support structure using modern CFC material, iii) building prototypes for test beams, and iv) integration in the overall detector concepts. Strongly motivated by the write up of the LOI due in March 2009, progress was achieved this last year in designing the detailed Silicon tracking systems for the ILC detector concepts. Especially challenging is the integration of a combined gaseous and Silicon tracking system as in the ILD case. The detailed construction of each of the Silicon components in the innermost and outermost parts both in the barrel and the Endcap regions are developed. SiLC is studying in details the case of an all Silicon tracking system. In parallel, detailed simulation GEANT 4 based (MOKKA and ILCROOT) complete the work performed with fast but already performing simulation (LiCToy and SGV) [7].

## 3 Test beams in 2008

Three test set-ups were installed, two at CERN with the standalone Si tracking system and the tests of the HPK test structures, and one at DESY with the LPTPC test [2] and (Fig.3).

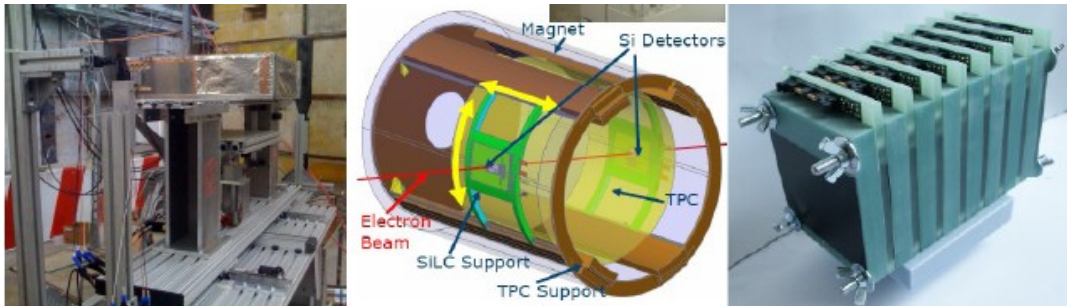

**Figure 3:** Standalone Si prototype at CERN test beam (left); schematic view of LPTPC DESY test beam (middle); HPK sensor test structures system for SPS-CERN test (right)

## 4 SiLC R&D Perspectives

SiLC has achieved a very success program of work since LCWS07 with new results presented at LCWS08. The tracking issues of the various ILC detector concepts are addressed with the ILD Si tracking combined with TPC, collaborative contacts with SiD and new collaboration with the 4$^{th}$ concept tracking system. It is pursued in synergy with the LHC and extends to the case of the CLIC machine. It starts from the present state of the art but is largely opened to the novel technologies, in close collaboration with the most relevant high tech industries.